\documentclass[conference]{IEEEtran}
\usepackage{cite}

\ifCLASSINFOpdf
   \usepackage[pdftex]{graphicx}
   \graphicspath{{./pdf/}{./jpeg}/}
   \DeclareGraphicsExtensions{.pdf,.jpeg,.png}
\else
\fi
\usepackage[cmex10]{amsmath}
\usepackage{array}

\ifCLASSOPTIONcompsoc
  \usepackage[caption=false,font=normalsize,labelfont=sf,textfont=sf]{subfig}
\else
  \usepackage[caption=false,font=footnotesize]{subfig}
\fi
\usepackage{siunitx}
\usepackage{bm}
\usepackage{bbm}
\usepackage{amssymb}

\usepackage{tikz}
\usetikzlibrary{decorations.pathreplacing}
\usetikzlibrary{shapes,arrows,fit,positioning}
\usetikzlibrary{plotmarks}
\usetikzlibrary{patterns}
\tikzset{>=latex}
\tikzstyle{block} = [draw, fill=white, rectangle,minimum height=3em, minimum width=5.2em]	
\tikzstyle{block_rot} = [draw, fill=white, rectangle,minimum height=5.2em, minimum width=3em]
\tikzstyle{coord} = [coordinate]
\tikzstyle{sum} = [draw, fill=white, circle, node distance=1em,path picture={\draw[black](path picture bounding box.south) -- (path picture bounding box.north) (path picture bounding box.west) -- (path picture bounding box.east);}]

\usepackage{pgfplots}
\pgfplotsset{compat=newest}
\newlength\figureheight	
\newlength\figurewidth
\newcommand{\columnplot}{\setlength\figureheight{0.3\textwidth} \setlength\figurewidth{0.4\textwidth}}	
\newcommand{\multiplot}{\setlength\figureheight{0.13\textwidth} \setlength\figurewidth{0.13\textwidth}}	
\newcommand{\pageplot}{\setlength\figureheight{0.23\textwidth} \setlength\figurewidth{0.23\textwidth}}	

\begin{document}
%
\title{Information Rates for Faster-Than-Nyquist Signaling with 1-Bit Quantization and Oversampling at the Receiver}

\author{\IEEEauthorblockN{Tim H{\"a}lsig, Lukas Landau, and Gerhard Fettweis}
\IEEEauthorblockA{Vodafone Chair Mobile Communications Systems\\
Technische Universit\"{a}t Dresden, 01062 Dresden, Germany\\
Email: tim.haelsig@ifn.et.tu-dresden.de, \{lukas.landau, fettweis\}@tu-dresden.de}}



%


\maketitle

\begin{abstract}

Oversampling combined with low quantization resolutions has been shown to be a viable option when aiming for energy efficiency in multigigabit/s communications systems. This work considers the case of 1-bit quantization combined with oversampling and shows how the performance of such a system can be improved by using matched pulse shaping filters and faster than Nyquist signaling. The channel is considered with additive Gaussian noise and the performance of the system is evaluated in terms of achievable information rate under symbol-by-symbol detection.

\end{abstract}

\let\thefootnote\relax\footnotetext{This work has been supported in part by the German Research Foundation in the framework of the Collaborative Research Center 912 "Highly Adaptive Energy-Efficient Computing" and by the European Social Fund in the framework of the Young Investigators Group "3D Chip-Stack Intraconnects".}

%
\IEEEpeerreviewmaketitle

\section{Introduction}
Increasing data rates of todays communications systems and thus increasing signaling bandwidths, demand for higher sampling rates at the receiver. One bottleneck, with increasing bandwidth and sampling rate, is the analog-to-digital conversion (ADC) in the receiver. Contemporary converters with fine grained quantization and high sampling rates tend to have a high power consumption \cite{le2005analog}. This can be counteracted by limiting the resolution to fewer bits (1-3 bits). The extreme case of one bit quantization in combination with oversampling with respect to symbol rate has been shown to be applicable \cite{cvetkovicsinglebit}, using a certain dithering signal. It has been shown that in many scenarios an even simpler architecture also provides a remarkable performance \cite{Shamai1994}, \cite{krone2012communications}, \cite{La12} and is thus under further investigation in this work. 

Over the last decade some investigations in regard to coarse quantization have been carried out.
In \cite{singh2009limits} the optimal input distributions with average power constraint, which maximize information rates depending on the signal-to-noise ratio (SNR), have been evaluated for different quantization resolutions.
It has been shown that for 1-bit quantization and additive white Gaussian noise (AWGN) channels, binary phase shift keying (BPSK) is optimal and achieves channel capacity when sampling at symbol rate. In \cite{krone2010achievable} 4-QAM and 16-QAM were shown to produce good results, with the latter yielding rates above two bits per symbol for a certain SNR range when employing oversampling. It was also shown that in those cases higher order modulation schemes (i.e. 16-QAM) gain more information rate from oversampling than lower ones (i.e. 4-QAM). In \cite{krone2012communications} it was shown that it is possible to benefit from intersymbol interference (ISI) when using 1-bit quantization and oversampling, where the interference appears like dithering. Furthermore the work in \cite{La12} indicates significantly increased information rates when considering 1-bit oversampling with ISI in combination with sequence estimation receivers.

In contrast to the existing literature \cite{krone2012communications}, this work also considers signaling rates above Nyquist rate. This approach is investigated as it is promising to increase the spectral efficiency. Furthermore different pulse shapes than previously have been used. It turned out that, especially when using root-raised-cosine (RRC) filters, information rates can be increased by employing signaling rates above Nyquist rate.

The paper is organized as follows: Section \ref{sec:model} gives an overview of the considered system model. In Section \ref{sec:rates} the calculation of the information rate for the system is illustrated. Section \ref{sec:nyquist} explains the approach of signaling rates above Nyquist rate \cite{mazo1975faster}. The numerical performance results and their implications are examined in \ref{sec:results}. The final Section \ref{sec:conc} concludes the paper with a brief summary of the addressed topics. 

In this paper bold letters describe vectors and bold capital letters define matrices such as $\bm{y}_k$ and $\bm{H}$, furthermore sequences of symbols are denoted as $x_{k-L/2}^{k+L/2}=\left[x_{k-L/2},\ldots,x_{k+L/2}\right]$.

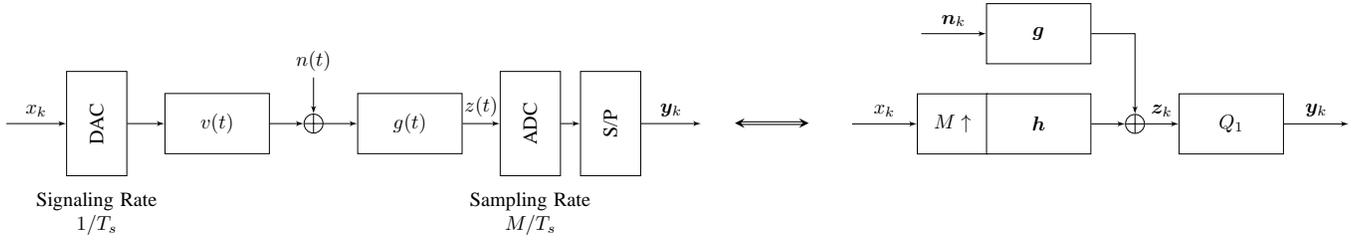
\begin{figure*}[!t]
\centering
\resizebox{0.52\textwidth}{!}{
	\begin{tikzpicture}[auto, node distance=4em,>=latex']
    
		\node [coord,node distance=3.5em] (input) {};
    \node [block_rot,node distance=4.5em, right of=input] (dac) {\rotatebox{90}{DAC}};
    \node [block, right of=dac,node distance=6em] (vt) {$v(t)$};

    \node [sum,right of=vt,node distance=4.8em] (sum) {};
    \node [block, right of=sum,node distance=4.8em] (gt) {$g(t)$};
    \node [coord,label={$n(t)$},above of=sum,node distance=2.3em] (noise) {};
    \node [block_rot, right of=gt,node distance=6em] (adc) {\rotatebox{90}{ADC}};
    \node [block_rot, right of=adc,node distance=4em] (sp) {\rotatebox{90}{S/P}};
    \node [coord,node distance=4.5em, right of=sp] (output) {};
    
		\node [coord,below of=dac,node distance=4.7em,label={Signaling Rate}] (signaling1) {};
		\node [coord,below of=dac,node distance=5.9em,label={$1/T_s$}] (signaling2) {};
    \node [coord,below of=adc,node distance=4.7em,label={Sampling Rate}] (sampling1) {};
		\node [coord,below of=adc,node distance=5.9em,label={$M/T_s$}] (sampling2) {};

    \draw [draw,->] (input) -- node {$x_k$} (dac);
    \draw [->] (dac) -- node {} (vt);
    \draw [->] (vt) -- node {}(sum);
    \draw [->] (noise) -- node {}(sum);
    \draw [->] (sum) -- node {}(gt);
    \draw [->] (gt) -- node {$z(t)$}(adc);	
    \draw [->] (adc) -- node {}(sp);
    \draw [->] (sp) -- node {$\bm{y}_k$}(output);

\end{tikzpicture}}
\raisebox{4.15em}{\begin{tikzpicture}[auto, node distance=4em,>=latex']

    \node[](arrow_start) {};
		\node[right of=arrow_start,node distance=3.5em] (arrow_end) {};
		\draw [<->,double,>=stealth] (arrow_start) -- node {} (arrow_end);
		
\end{tikzpicture}}
\raisebox{3.3em}{\resizebox{0.38\textwidth}{!}{
	\begin{tikzpicture}[auto, node distance=4em,>=latex']
    \node [coord,right of=arrow_end,node distance=2.4em] (input_disc) {};
    
    \node [block, right of=input_disc,node distance=5em,minimum width=3.5em] (upsampling) {$M\uparrow$};
    \node [block, right of=upsampling,node distance=4.3em] (h) {$\bm{h}$};

    \node [block, above of=h,node distance=4.5em] (g) {$\bm{g}$};
    \node [coord, left of=g,node distance=5.85em] (noise_disc) {};
    
    \node [sum,right of=h,node distance=4.8em] (sum) {};
    
    \node [block, right of=sum,node distance=4.8em] (1bit) {$Q_1$};

    \node [coord,node distance=5.85em, right of=1bit] (output_disc) {};

    \draw [draw,->] (input_disc) -- node {$x_k$} (upsampling);
    \draw [->] (h) -- node {}(sum);
    \draw [->] (noise_disc) -- node {$\bm{n}_k$}(g);	
    \draw [->] (g) -| node {}(sum);
    \draw [->] (sum) -- node {$\bm{z}_k$}(1bit);
    \draw [->] (1bit) -- node {$\bm{y}_k$}(output_disc);

\end{tikzpicture}
		}}
		
\caption{Continous and corresponding discrete system model of a communications system that transmits complex valued data over an AWGN channel with matched filters and uses 1-bit conversion and oversampling.}
\label{fig:model}
\end{figure*}

\section{System Model} \label{sec:model}
The received signal $z(t)$ of a communications system with complex input symbols drawn from a finite set $x_k \in \mathbb{X}$ (symbol rate $1/T_s$), two matched filters $v(t)$, $g(t)$ and additive white Gaussian noise $n(t)$ is given by
\begin{align}
z(t)=&\left(\sum_{k=-\infty}^{\infty}{x_k \delta(t-kT_s)} \ast v(t) + n(t) \right)\ast g(t) \text{,}
\end{align}
where both filters have unit energy. The combined channel waveform is denoted as $h(t)=v(t) \ast g(t)$.
The discrete quantized $M$-fold oversampled signal is denoted as
\begin{align}
y_{k,m} = Q_1 \left[ z\left(kT_s - \frac{T_s}{2}\left(\frac{M+1}{M}\right)+\frac{mT_s}{M}\right) \right] \text{,}
\end{align}
where $Q_1[\cdot]$ describes the 1-bit quantization operator 
\begin{align}
y_{k,m} = Q_1 \left[ \bm{z}_{k,m}\right] = \begin{cases}  \phantom{-} 1 & \mbox{for } \bm{z}_{k,m}\geq 0 \\
-1 & \mbox{otherwise}
\end{cases}
\end{align}
and $m=1\ldots M$ is the running index sampling $M$ times per symbol duration $T_s$. The stacked $M$ samples per symbol duration are denoted in vector notation as $\bm{y}_k=\left[y_{k,1},\ldots,y_{k,M}\right]^T$.  

In the corresponding discrete system model, which can be seen in Figure \ref{fig:model}, oversampling is realized by upsampling the input symbols. It is considered that the filter $h$ has the length of $L+1$ symbols and the filter $g$ has the length of $N$ symbols ($L$ and $N$ are even). As the receive filter has unit energy the noise variance at the output of the filter is equivalent to the noise power density $\sigma_{n}^2$. This is modeled equivalently with i.i.d noise samples $\bm{n}=\left[n_{k,1},\ldots,n_{k,M}\right]^T$ of variance $\sigma_{n}^2$. The system follows with
\begin{align}
\bm{y}_k = Q_1 \left[ \bm{H} \bm{U} x_{k-L/2}^{k+L/2} + \bm{G} \bm{n}_{k-N/2}^{k+N/2} \right] \text{,}
\end{align}
where $\bm{U}$ is the upsampling matrix of the dimension $(L+2)M-1 \times L+1$. Its elements are given by
\begin{align}
U_{i,j} = \begin{cases}
1 & \mbox{for } i=j\cdot M  \\
0 & \mbox{otherwise.}
\end{cases}
\end{align}
The filter matrices $\bm{H}$ with dimension $M \times (L+2)M-1$ and $\bm{G}$ with dimension $M \times (N+1)M$ have a Toeplitz structure as follows
\begin{align}
\bm{H}=
\begin{pmatrix}
\left[ \ \bm{h}_{\text{r}}^T \ \right]  \ 0 \cdots \ \ \ 0 \\
0 \ \left[ \ \bm{h}_{\text{r}}^T \ \right] \ 0 \cdots 0 \\
\ddots  \ddots \ddots  \\
0 \cdots \ \ \ 0 \ \left[ \ \boldsymbol{h}_{\text{r}}^T \ \right]
\end{pmatrix} \textrm{, }
\bm{G}=\begin{pmatrix}
\left[ \ \bm{g}_{\text{r}}^T \ \right]  \ 0 \cdots \ \ \ 0 & 0 \\
0 \ \left[ \ \bm{g}_{\text{r}}^T \ \right] \ 0 \cdots 0 & 0 \\
\ddots  \ddots \ddots  & \vdots \\
0 \cdots \ \ \  0 \ \left[ \ \bm{g}_{\text{r}}^T \ \right] & 0 
\end{pmatrix} \textrm{.} 
\end{align}
They are signified by the reversed vector of the corresponding filter, i.e. $\bm{h}_\text{r} = \left[h_{(L+1)M},\ldots,h_{1}\right]^T$respectively $\bm{g}_\text{r} = \left[g_{N\cdot M},\ldots,g_{1}\right]^T$, followed by zeros such that they match the defined dimensions.

\section{Information Rate} \label{sec:rates}
As a measure for the achievable throughput a lower bound on the information rate (see the Appendix) is considered. Indeed the lower bound for i.i.d. input given by 
\begin{align}
\lim_{n \to \infty} \frac{1}{n} I(X^n;\bm{Y}^n) \geq I(X_k;\bm{Y}_k) 
\end{align}
corresponds to the information rate that treats the channel as an equivalent discrete memoryless channel (DMC). Albeit the present channel has memory which could be exploited, this work considers a simple receiver architecture which employs symbol-by-symbol detection. To compute this information rate, the mutual information of a DMC is used which is given by
\begin{align}           
I(X_k;\bm{Y}_k) & = \sum_{\bm{y}_k\in \mathbb{Y}}\sum_{x_k\in \mathbb{X}}P(\bm{y}_k|x_k)P(x_k)\log_2\left(\frac{P(\bm{y}_k|x_k)}{P(\bm{y}_k)}\right) \text{,}
\end{align}
where $P(x_k)$ and $P(\bm{y}_k)$ are the probability distributions of the sent and received symbols respectively.

Caused by the waveform and oversampling each transmitted symbol interferes with every sample taken from previous and future symbols, with exception of sampling at Nyquist rate (no oversampling) and using Nyquist pulses. Therefore the transition probability is as follows
\begin{align}
P(\bm{y}_k|x_k) & = \sum_{x_{k-L/2}^{k+L/2}}P\left(x_{k-L/2}^{k+L/2}\right)P\left(\bm{y}_k \big| x_{k-L/2}^{k+L/2}\right) \text{,}
\end{align}
where $L$ is the number of neighboring symbols that influence the current oversampling observation $\bm{y}_k$.

The unquantized received signal $\bm{z}_k$ is characterized by the additive white Gaussian noise with the specific mean and variance determined by the filters. The mean is calculated by
\begin{align}
\bm{\mu}_k & = \bm{H}\bm{U} x_{k-L/2}^{k+L/2}
\end{align}
and the noise covariance matrix is given by 
\begin{align}
\bm{R} & =  E\left[\bm{G} \bm{n}_{k-N/2}^{k+N/2} \left(\bm{n}_{k-N/2}^{k+N/2}\right)^{\mathrm{H}} \bm{G}^{\mathrm{H}} \right] = \sigma_{n}^{2} \bm{G} \bm{G}^{\mathrm{H}} \text{.}
\end{align}
The probability density function is then given by the multivariate complex Gaussian distribution, written as
\begin{align}
p& \left(\bm{z}_k  \big| x_{k-L/2}^{k+L/2} \right) = \notag \\
	& \frac{1}{\pi^M \left|\bm{R}\right|} \cdot \exp\left(-(\bm{z}_k-\bm{\mu}_k)^{\mathrm{H}} \bm{R}^{-1} (\bm{z}_k-\bm{\mu}_k)\right) \label{eq:pdf}
\text{.}
\end{align}
The final transition probability can be found by integrating \eqref{eq:pdf} over the quantization interval
\begin{align}
P\left(\bm{y}_k \big| x_{k-L/2}^{k+L/2} \right) & = \int\limits_{\bm{z}_k\in \mathbb{Y}_{k}} p\left(\bm{z}_k \big| x_{k-L/2}^{k+L/2}\right) \,\operatorname{d}\bm{z}_k \text{,}
\end{align}
where $\mathbb{Y}_{k}=\left\{\bm{z}_k \vert Q_1[\bm{z}_k]=\bm{y}_k\right\}$, which corresponds to integration limits of minus infinity and zero or zero and infinity.

\section{Signaling Rates above Nyquist Rate} \label{sec:nyquist}
The following sections rely on the fact that the bandwidth $B$ is only determined by the pulse shaping filters $v(t)$ and $g(t)$ included in $h(t)$. When considering Nyquist pulses the signaling rate is conventionally chosen as $1/T_s=B$. Following the approach in \cite{mazo1975faster} and \cite{Anderson2013}, cases with faster than Nyquist signaling rate are also considered, which corresponds to
\begin{equation}
B\cdot T_{s} \leq 1 \text{.} \label{eq:fasternyquist}
\end{equation}
Applying faster signaling rates while keeping the bandwidth constant increases the amount of intersymbol interference (ISI) between neighboring symbols. On one hand this leads to a decreased system performance, however, due to the increased signaling rate this effect can be compensated to some extent. Looking at the overall system performance it is assumed that a beneficial tradeoff between mutual information per channel use and signaling rate exists.

This tradeoff corresponds to an increased spectral efficiency which might be implemented with low complexity, suitable for multigigabit/s communications. As for oversampling receivers interference is already present in general it is promising to employ faster signaling rates. In the following section those effects are investigated by carrying out numerical simulations with oversampling and 1-bit conversion.

\section{Numerical Results} \label{sec:results}
The numerical results have been obtained by considering i.i.d. symbols from the 4-QAM respectively 16-QAM alphabet. The system is assumed to be coherent.

Two pulse shaping filter types have been applied, namely the Gaussian shaped filter and the Root-Raised-Cosine filter, which are commonly used in various digital communications systems. Such filters can be designed variable and energy efficient in analog domain as it is proposed in \cite{thakkar2012cmos} and \cite{zhao201017}. The Gaussian and RRC filters are characterized by parameters that determine shape and bandwidth. The Gaussian filter is given by 
\begin{equation}
v(t)=\frac{\sqrt{2\pi}}{\sqrt{\ln\left(2\right)}}\frac{B_{\text{3dB}}T_{s}}{2T_{s}}\exp\left[-\left(\frac{\sqrt{2}\pi}{\sqrt{\ln\left(2\right)}}\frac{B_{\text{3dB}}T_{s}}{2T_{s}}t\right)^{2}\right] \label{eq:gauss} \text{,}
\end{equation}
where the product $B_{\text{3dB}}T_{s}$ is the design parameter controlling the pulse shape.
The RRC filter is determined by
\small
\begin{equation}
v(t)=\begin{cases}
1-\beta+4\frac{\beta}{\pi} & t=0\\
\frac{\beta}{\sqrt{2}}\left[\left(1+\frac{2}{\pi}\right)\sin\left(\frac{\pi}{4\beta}\right)+\left(1-\frac{2}{\pi}\right)\cos\left(\frac{\pi}{4\beta}\right)\right] & t=\pm\frac{T_{x}}{4\beta}\\
\frac{\sin\left[\pi\frac{t}{T_{x}}\left(1-\beta\right)\right]+4\beta\frac{t}{T_{x}}\cos\left[\pi\frac{t}{T_{x}}\left(1+\beta\right)\right]}{\pi\frac{t}{T_{x}}\left[1-\left(4\beta\frac{t}{T_{x}}\right)^{2}\right]} & \text{,}
\end{cases}
\end{equation}
\normalsize
where $1/T_x$ determines the \SI{3}{\decibel} bandwidth and $\beta$ is the roll-off factor of the pulse. It is furthermore assumed that $v(t)=g(t)$ and $h(t)=v(t)\ast g(t)$. As those filters have an infinite impulse response, an approximation with finite length of $L+1=9$ symbol durations has been applied to reduce computational complexity, given by their discrete representations $\bm{h}$ and $\bm{g}$. 

All obtained results were computed using a Monte-Carlo-Simulation in order to estimate $P(\bm{y}|x)$ and $P(\bm{y})$. The symmetry of I and Q has been taken into account by simulating only one phase component to ease computation, which is sufficient since coherence is assumed.

\subsection{Comparison Gauss and Root-Raised-Cosine Filters at Conventional Signaling Rate} \label{sec:comp}
\begin{figure}[!htb]
\centering
\columnplot
%
%
%
\definecolor{mycolor1}{rgb}{0.50196,0.50196,0.50196}%
\definecolor{mycolor2}{rgb}{0.2,0.6,1}%
\begin{tikzpicture}

\begin{axis}[%
font=\small,
width=\figurewidth,
height=\figureheight,
scale only axis,
xmin=-5,
xmax=35,
xlabel={SNR in dB},
xmajorgrids,
ymin=0,
ymax=4.5,
ylabel={$I$ in bpcu},
ymajorgrids,
legend style={at={(0.01,0.99)},anchor=north west,draw=black,fill=white,legend cell align=left,font=\scriptsize} 
]
\addplot [color=mycolor1,dotted,line width=1.0pt]
  table[row sep=crcr]{
-5	0.33764626	\\
-3	0.559449283	\\
-1	0.831693803	\\
1	1.15775082	\\
3	1.54070686	\\
5	1.972649628	\\
7	2.440431353	\\
9	2.925494492	\\
11	3.383256079	\\
13	3.735075449	\\
15	3.927218027	\\
17	3.989957367	\\
19	3.999521951	\\
21	3.999995649	\\
23	3.999999997	\\
25	4	\\
27	4	\\
29	4	\\
31	4	\\
33	4	\\
35	4	\\
};
\addlegendentry{No Quantization};

\addplot [color=red,solid,line width=1.0pt,mark=triangle,mark options={solid,,rotate=180}]
  table[row sep=crcr]{
-5	0.3243751725	\\
-3	0.4805168114	\\
-1	0.6794727723	\\
1	0.9272741613	\\
3	1.1945207386	\\
5	1.4681819428	\\
7	1.7335295955	\\
9	1.9694178635	\\
11	2.2020886901	\\
13	2.3927484612	\\
15	2.5218081352	\\
17	2.5967283144	\\
19	2.624808626	\\
21	2.6346032422	\\
23	2.6362261825	\\
25	2.6355441996	\\
27	2.6345163129	\\
29	2.6306523443	\\
31	2.6297056155	\\
33	2.6267962066	\\
35	2.6248672128	\\
};
\addlegendentry{RRC $\beta=0.6$};

\addplot [color=red,solid,line width=1.0pt,mark=triangle,mark options={solid}]
  table[row sep=crcr]{													
-5	0.3281044406	\\
-3	0.4760377716	\\
-1	0.6718555403	\\
1	0.9173413509	\\
3	1.1923375196	\\
5	1.476824002	\\
7	1.7424989799	\\
9	1.9616098168	\\
11	2.174497054	\\
13	2.3328240591	\\
15	2.4398062406	\\
17	2.4980851596	\\
19	2.5345374172	\\
21	2.5550365314	\\
23	2.569392629	\\
25	2.5730821736	\\
27	2.5847896641	\\
29	2.5844344307	\\
31	2.5876360126	\\
33	2.5879593671	\\
35	2.5906321807	\\
};
\addlegendentry{RRC $\beta=0.3$};

%

\addplot [color=mycolor2,solid,line width=1.0pt,mark=o,mark options={solid}]
  table[row sep=crcr]{                                                                 
-5	0.304292171102338	\\
-3	0.449016568461503	\\
-1	0.644218229779298	\\
1	0.887545611273575	\\
3	1.15449857466286	\\
5	1.42338307082627	\\
7	1.66226160656355	\\
9	1.88456646715748	\\
11	2.08608713948116	\\
13	2.25645717630960	\\
15	2.36272724455840 	\\
17	2.40242375532511	\\
19	2.40760564117326	\\
21	2.41505089436186	\\
23	2.43014027679624	\\
25	2.45716763790228	\\
27	2.48517024533221	\\
29	2.50830827929673	\\
31	2.52255296743359	\\
33	2.52860730558913	\\
35	2.52999815800013	\\
};
\addlegendentry{Gauss $B_{\text{3dB}}T_s=1$};

\addplot [color=black,dashed]
  table[row sep=crcr]{
-5	0.263083358	\\
-3	0.39541635	\\
-1	0.579574124	\\
1	0.821492491	\\
3	1.116784319	\\
5	1.447454572	\\
7	1.783888063	\\
9	2.087548056	\\
11	2.306183547	\\
13	2.383439863	\\
15	2.323799003	\\
17	2.201754315	\\
19	2.088735924	\\
21	2.023980453	\\
23	2.003167386	\\
25	2.000139952	\\
27	2.000001112	\\
29	2.000000001	\\
31	2	\\
33	2	\\
35	2	\\
};
\addlegendentry{Rectangular};

\end{axis}
\end{tikzpicture}%
\caption{Information rate of 16-QAM with different matched filters, filter length $9$ symbols and oversampling factor $M=4$.}
\label{fig:pulsecomp}
\end{figure}
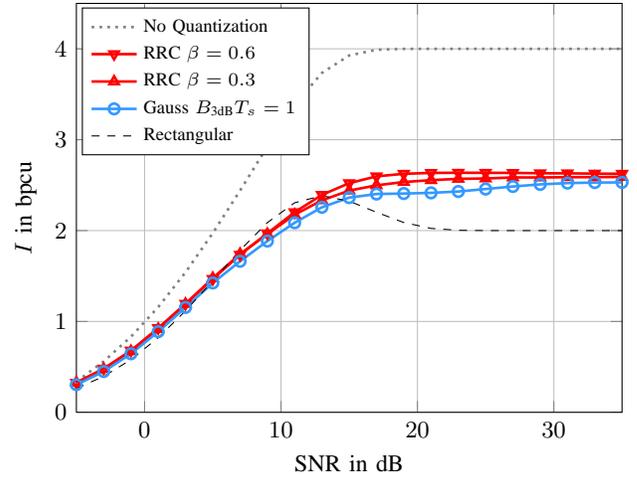

The simulation results show that both pulses can lead to an increased information rate compared to results as reached in \cite{krone2010achievable} (in Figure \ref{fig:pulsecomp} Rectangular). This gain can be explained by an intersymbol interference based dithering utilization. For RRC pulses this effect is stronger as compared to the Gaussian pulse. According to our experience with this scenario, the benefit of increasing the oversampling rate further only leads to marginal performance gains, which can be explained by a strong correlation of the individual samples.

The results shown in Figure \ref{fig:pulsecomp} provide an example of different parameters. 

Another benefit that can be seen is the stabilization of information rates at higher SNR values. This might be useful for receiver implementations that need to operate within a wide SNR range. 

\begin{figure*}[!t]
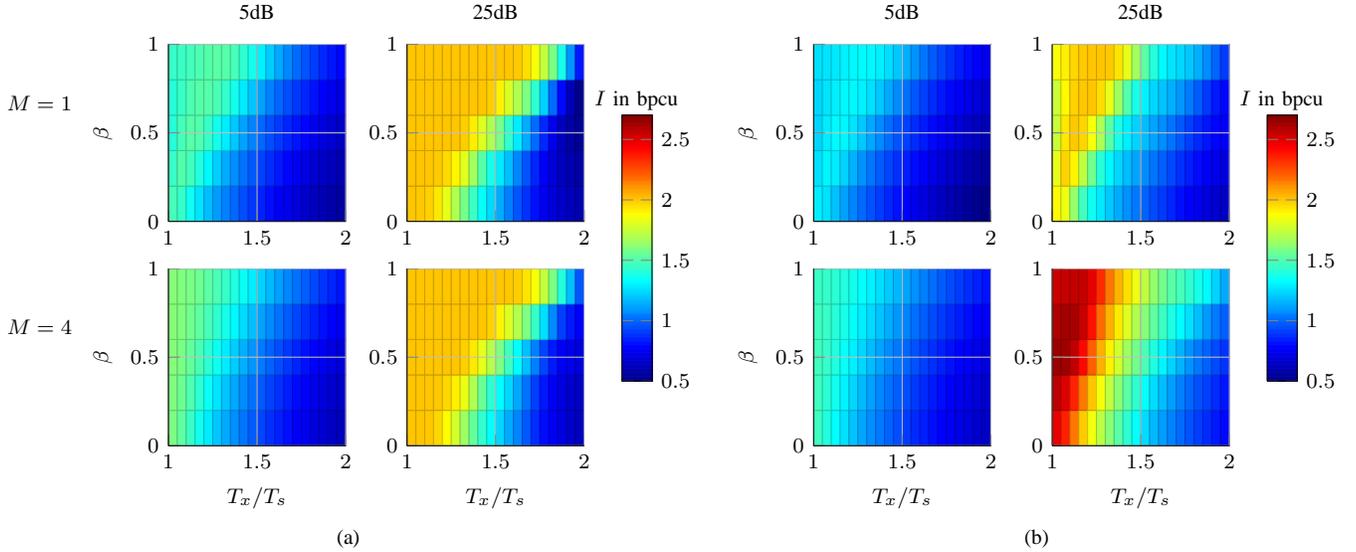

\centering
\subfloat[]{\multiplot\input{figures/4qamfasterNyquist.tikz}
\label{fig:4qamunnorm}}
\hfill
\subfloat[]{\multiplot\input{figures/16qamfasterNyquist.tikz}
\label{fig:16qamunnorm}}
\caption{Information rate results of \protect\subref{fig:4qamunnorm} 4-QAM and \protect\subref{fig:16qamunnorm} 16-QAM, at \SI{5}{\decibel} and \SI{25}{\decibel} for different roll-offs $\beta$ and different signaling rates $T_x/T_s$, oversampling $M=1$ and $M=4$, filter length $9$ symbols.}
\label{fig:mutinf}
\end{figure*}

\subsection{Tradeoff when employing faster signaling rates with RRC pulses}
In this section the performance gain that can be achieved by signaling faster than Nyquist is going to be considered. Although the achievable rate, when considering Gaussian pulses, is rather robust against \SI{3}{\decibel} bandwidth reduction, they are not considered because of their overall infinite bandwidth and the following investigations are only done with RRC pulses.

The idea is that higher signaling rates can be achieved by fixing the pulses \SI{3}{\decibel} bandwidth $B_{\text{3dB}}=1/T_x$ and reducing the symbol duration to values smaller than the inverse of the bandwidth $T_{x}>T_{s}$. Equivalently it is possible to to reduce the \SI{3}{\decibel} bandwidth of the system by fixing symbol duration $T_s$ while reducing bandwidth $1/T_x$. These two options are interchangeable since only the relation between \SI{3}{\decibel} bandwidth and symbol duration matters, $B_{\text{3dB}}\cdot T_s = T_s /T_x$. Conventional faster than Nyquist signaling is for RRC pulses achieved when
\begin{equation}
\frac{T_x}{T_s} > 1+\beta
\text{}
\end{equation}
holds. This method is especially promising for lower order modulation schemes and hence it was also employed with 4-QAM, which seemed feasible for the use with 1-bit quantization in previous investigations.

With roll-off factor $\beta$ and signaling parameter $T_{x}/T_{s}$ ($T_{x}/T_{s}=1$ is the conventional signaling case examined in Section \ref{sec:comp}) being variable, a three dimensional optimization problem emerges when searching for maximum information rate at any given SNR. Note that the parameter $T_{x}/T_{s}$ is the inverse of the parameter $B_\text{3dB}\cdot T_s$ and thus only values greater than $1$ are of interest.

From the numerical results, examples shown in Figure \ref{fig:mutinf} at  \SI{5}{\decibel} and \SI{25}{\decibel}, different observations can be made. First in most of the cases an increase of the roll-off factor improves the information rate. The second result is that the full rate of $2$ bits per channel use (bpcu), when using 4-QAM, is achievable for a number of scenarios with $B_\text{3dB}\cdot T_s<1$. Such a range exists for 16-QAM as well but is substantially smaller. The maximum achievable rate can, however, be higher than $2$ bpcu ($M=4$), but is more strongly dependent on the SNR and is hence also more susceptible to the additional interference brought by the faster than signaling approach. Oversampling is especially beneficial in the lower SNR regime and for achieving more than $2$ bpcu when using 16-QAM.
Note that results for $T_{x}/T_{s}<1$ provide similar results but would require more bandwidth and are therefore not desired for this concept.

Since faster signaling cases are applicable the achievable rate would be higher compared to systems signaling with $T_x=T_s$. Mutual information per channel use as considered before does not account for that performance increase. Therefore an information rate shall be considered that includes the gain of the increased signaling rates.

\begin{figure*}[!t]
\subfloat[]{\pageplot
%
%
\begin{tikzpicture}[font=\footnotesize,]

\begin{axis}[%
width=\figurewidth,
height=\figureheight,
view={0}{90},
scale only axis,
xmin=1,
xmax=2,
xlabel={$T_x/T_s$},
xmajorgrids,
ymin=0,
ymax=1,
ylabel={$\beta$},
ymajorgrids,
zmin=0,
zmax=4,
zmajorgrids,
name=plot1,
title={4-QAM},
axis x line*=bottom,
axis y line*=left,
axis z line*=left,
point meta min=1.2,
point meta max=3.6,
]

\addplot3[%
surf,
shader=faceted,
draw=black,
colormap/jet,
mesh/rows=21]
table[row sep=crcr,header=false] {
1	0	1.999997772	\\
1	0.2	1.999988674	\\
1	0.4	1.999988035	\\
1	0.6	1.999996875	\\
1	0.8	1.999943558	\\
1	1	1.999990718	\\
1.05	0	2.0999989	\\
1.05	0.2	2.099951406	\\
1.05	0.4	2.099958593	\\
1.05	0.6	2.099982538	\\
1.05	0.8	2.09998473	\\
1.05	1	2.099992627	\\
1.1	0	2.19994655	\\
1.1	0.2	2.19999295	\\
1.1	0.4	2.199978919	\\
1.1	0.6	2.199982316	\\
1.1	0.8	2.199974547	\\
1.1	1	2.199975716	\\
1.15	0	2.271280798	\\
1.15	0.2	2.299978446	\\
1.15	0.4	2.299972055	\\
1.15	0.6	2.299987071	\\
1.15	0.8	2.299975652	\\
1.15	1	2.29997569	\\
1.2	0	2.284586179	\\
1.2	0.2	2.368996124	\\
1.2	0.4	2.399993686	\\
1.2	0.6	2.399985028	\\
1.2	0.8	2.399944553	\\
1.2	1	2.399976276	\\
1.25	0	2.230979736	\\
1.25	0.2	2.348245064	\\
1.25	0.4	2.499981477	\\
1.25	0.6	2.499990286	\\
1.25	0.8	2.49997226	\\
1.25	1	2.499989755	\\
1.3	0	2.204443674	\\
1.3	0.2	2.300106404	\\
1.3	0.4	2.592700346	\\
1.3	0.6	2.599979498	\\
1.3	0.8	2.599998001	\\
1.3	1	2.599987351	\\
1.35	0	2.124036679	\\
1.35	0.2	2.211803065	\\
1.35	0.4	2.600781402	\\
1.35	0.6	2.699758546	\\
1.35	0.8	2.699963559	\\
1.35	1	2.699994042	\\
1.4	0	2.047485296	\\
1.4	0.2	2.104944244	\\
1.4	0.4	2.470319382	\\
1.4	0.6	2.799521287	\\
1.4	0.8	2.799928516	\\
1.4	1	2.799996244	\\
1.45	0	1.969249774	\\
1.45	0.2	2.044037475	\\
1.45	0.4	2.242679155	\\
1.45	0.6	2.882701885	\\
1.45	0.8	2.899891066	\\
1.45	1	2.899945869	\\
1.5	0	1.885716843	\\
1.5	0.2	1.968401731	\\
1.5	0.4	2.138700304	\\
1.5	0.6	2.851321411	\\
1.5	0.8	2.999975602	\\
1.5	1	2.999983549	\\
1.55	0	1.812653175	\\
1.55	0.2	1.880698539	\\
1.55	0.4	2.110654001	\\
1.55	0.6	2.637056939	\\
1.55	0.8	3.098971265	\\
1.55	1	3.099984039	\\
1.6	0	1.711787399	\\
1.6	0.2	1.799309142	\\
1.6	0.4	2.008167824	\\
1.6	0.6	2.440254079	\\
1.6	0.8	3.178191105	\\
1.6	1	3.19998122	\\
1.65	0	1.630218354	\\
1.65	0.2	1.74461559	\\
1.65	0.4	1.837124648	\\
1.65	0.6	2.319193199	\\
1.65	0.8	3.19647797	\\
1.65	1	3.299701864	\\
1.7	0	1.528234482	\\
1.7	0.2	1.569347773	\\
1.7	0.4	1.693347239	\\
1.7	0.6	2.121769333	\\
1.7	0.8	3.076370236	\\
1.7	1	3.399682498	\\
1.75	0	1.44515476	\\
1.75	0.2	1.459915771	\\
1.75	0.4	1.538550417	\\
1.75	0.6	1.886307041	\\
1.75	0.8	2.849364047	\\
1.75	1	3.4869809	\\
1.8	0	1.378663308	\\
1.8	0.2	1.392754785	\\
1.8	0.4	1.451855951	\\
1.8	0.6	1.66876361	\\
1.8	0.8	2.524063066	\\
1.8	1	3.538959149	\\
1.85	0	1.352371798	\\
1.85	0.2	1.35648669	\\
1.85	0.4	1.353687981	\\
1.85	0.6	1.560316153	\\
1.85	0.8	2.087135509	\\
1.85	1	3.479897122	\\
1.9	0	1.299864943	\\
1.9	0.2	1.297086152	\\
1.9	0.4	1.310458953	\\
1.9	0.6	1.418059147	\\
1.9	0.8	1.760221289	\\
1.9	1	3.188764312	\\
1.95	0	1.248072713	\\
1.95	0.2	1.25664404	\\
1.95	0.4	1.277794823	\\
1.95	0.6	1.608602099	\\
1.95	0.8	1.580988854	\\
1.95	1	2.594681783	\\
2	0	1.238748039	\\
2	0.2	1.273567524	\\
2	0.4	1.288533291	\\
2	0.6	1.325523831	\\
2	0.8	1.501523755	\\
2	1	1.920089672	\\
};
\addplot[thick] coordinates{(1,0) (2,1)};
\end{axis}

\begin{axis}[%
width=\figurewidth,
height=\figureheight,
view={0}{90},
scale only axis,
xmin=1,
xmax=2,
xlabel={$T_x/T_s$},
xmajorgrids,
ymin=0,
ymax=1,
ylabel={},	
ymajorgrids,
zmin=0,
zmax=4,
zmajorgrids,
name=plot2,
at=(plot1.right of south east),
anchor=left of south west,
title={16-QAM},
axis x line*=bottom,
axis y line*=left,
axis z line*=left,
point meta min=1.2,
point meta max=3.6,
]

\addplot3[%
surf,
shader=faceted,
draw=black,
colormap/jet,
mesh/rows=21]
table[row sep=crcr,header=false] {
1	0	2.471134	\\
1	0.2	2.532824096	\\
1	0.4	2.645386322	\\
1	0.6	2.630667741	\\
1	0.8	2.536194284	\\
1	1	2.518477646	\\
1.05	0	2.533795	\\
1.05	0.2	2.60494491	\\
1.05	0.4	2.75038642	\\
1.05	0.6	2.80242784	\\
1.05	0.8	2.685060742	\\
1.05	1	2.655451141	\\
1.1	0	2.434827	\\
1.1	0.2	2.566777023	\\
1.1	0.4	2.816369444	\\
1.1	0.6	2.950442776	\\
1.1	0.8	2.834432329	\\
1.1	1	2.778826196	\\
1.15	0	2.307293	\\
1.15	0.2	2.442352111	\\
1.15	0.4	2.759774307	\\
1.15	0.6	3.005449825	\\
1.15	0.8	2.976237835	\\
1.15	1	2.915282428	\\
1.2	0	2.233613	\\
1.2	0.2	2.350190335	\\
1.2	0.4	2.634642289	\\
1.2	0.6	2.936037634	\\
1.2	0.8	3.052596627	\\
1.2	1	3.043597489	\\
1.25	0	2.186255	\\
1.25	0.2	2.287047983	\\
1.25	0.4	2.523114752	\\
1.25	0.6	2.795513377	\\
1.25	0.8	2.931967219	\\
1.25	1	3.125645952	\\
1.3	0	2.134679	\\
1.3	0.2	2.209250833	\\
1.3	0.4	2.449051589	\\
1.3	0.6	2.663397636	\\
1.3	0.8	2.803654223	\\
1.3	1	3.031234163	\\
1.35	0	2.077902	\\
1.35	0.2	2.166608292	\\
1.35	0.4	2.376420857	\\
1.35	0.6	2.597200118	\\
1.35	0.8	2.758187529	\\
1.35	1	2.891310961	\\
1.4	0	2.015417	\\
1.4	0.2	2.091461958	\\
1.4	0.4	2.307031675	\\
1.4	0.6	2.518511051	\\
1.4	0.8	2.690336672	\\
1.4	1	2.884464843	\\
1.45	0	1.954207	\\
1.45	0.2	2.042463447	\\
1.45	0.4	2.234886979	\\
1.45	0.6	2.463369276	\\
1.45	0.8	2.610342533	\\
1.45	1	2.868302327	\\
1.5	0	1.897953	\\
1.5	0.2	1.968275659	\\
1.5	0.4	2.170035447	\\
1.5	0.6	2.415191641	\\
1.5	0.8	2.565669929	\\
1.5	1	2.797128641	\\
1.55	0	1.837224	\\
1.55	0.2	1.899179383	\\
1.55	0.4	2.106230217	\\
1.55	0.6	2.363734329	\\
1.55	0.8	2.520287935	\\
1.55	1	2.707166958	\\
1.6	0	1.790366	\\
1.6	0.2	1.863377865	\\
1.6	0.4	2.011145833	\\
1.6	0.6	2.290619923	\\
1.6	0.8	2.495382095	\\
1.6	1	2.638421847	\\
1.65	0	1.753706	\\
1.65	0.2	1.83769532	\\
1.65	0.4	1.967879658	\\
1.65	0.6	2.223967666	\\
1.65	0.8	2.463143113	\\
1.65	1	2.623889883	\\
1.7	0	1.724879	\\
1.7	0.2	1.770252957	\\
1.7	0.4	1.913003178	\\
1.7	0.6	2.149610396	\\
1.7	0.8	2.450143485	\\
1.7	1	2.63360469	\\
1.75	0	1.693707	\\
1.75	0.2	1.73345302	\\
1.75	0.4	1.860174783	\\
1.75	0.6	2.082818079	\\
1.75	0.8	2.444492257	\\
1.75	1	2.640178219	\\
1.8	0	1.655072	\\
1.8	0.2	1.678652692	\\
1.8	0.4	1.809516392	\\
1.8	0.6	2.013367936	\\
1.8	0.8	2.395006043	\\
1.8	1	2.647442641	\\
1.85	0	1.626859	\\
1.85	0.2	1.660138026	\\
1.85	0.4	1.761122437	\\
1.85	0.6	1.959904978	\\
1.85	0.8	2.289527809	\\
1.85	1	2.702478194	\\
1.9	0	1.587455	\\
1.9	0.2	1.636843692	\\
1.9	0.4	1.76578375	\\
1.9	0.6	1.90705333	\\
1.9	0.8	2.170519916	\\
1.9	1	2.697090021	\\
1.95	0	1.565438	\\
1.95	0.2	1.573348168	\\
1.95	0.4	1.680031168	\\
1.95	0.6	1.848484637	\\
1.95	0.8	2.064168196	\\
1.95	1	2.559012433	\\
2	0	1.526826	\\
2	0.2	1.571155724	\\
2	0.4	1.648845963	\\
2	0.6	1.811509534	\\
2	0.8	2.019125668	\\
2	1	2.382164354	\\
};

\addplot[thick] coordinates{(1,0) (2,1)};
\end{axis}

\begin{axis}[
    hide axis,
    scale only axis,
    height=0pt,
    width=0pt,
    colormap/jet,
		colorbar,
		name=bar1,
		at=(plot2.right of north east),
		yshift=-0.1\figureheight,
		xshift=2mm,
    point meta min=1.2,
    point meta max=3.6,
    colorbar style={
				width=4mm,
				height=0.8\figureheight,
    }]
\end{axis}

\node[xshift=5mm,yshift=4mm,align=center] at (bar1.north) {$I_\text{FTN}$ in\\ bits per $T_x$};

\end{tikzpicture}%
\label{fig:combinednorm}}
\subfloat[]{\pageplot
%
%
\begin{tikzpicture}[font=\footnotesize,]

\begin{axis}[%
width=\figurewidth,
height=\figureheight,
scale only axis,
xmin=1,
xmax=2,
xlabel={$T_x/T_s$},
ymin=0,
ymax=1,
ylabel={$\beta$},
name=plot1,
axis x line*=bottom,
axis y line*=left,
]

\addplot[pattern=crosshatch] coordinates{(0,0) (0,1) (1.4,1) (1.4,0.8) (1.35,0.8) (1.35,0.6) (1.3,0.6) (1.3,0.4) (1.2,0.4) (1.2,0.2) (1.55,0.2) (1.55,0.4) (1.6,0.4) (1.6,0.6) (1.7,0.6) (1.7,0.8) (1.85,0.8) (1.85,1) (2,1) (2,0) (0,0)};

\addplot[draw=black,shader=faceted, pattern=dots] coordinates{(1.2,0.2) (1.2,0.4) (1.3,0.4) (1.3,0.6) (1.35,0.6) (1.35,0.8) (1.4,0.8) (1.4,1) (1.85,1) (1.85,0.8) (1.7,0.8) (1.7,0.6) (1.6,0.6) (1.6,0.4) (1.55,0.4) (1.55,0.2) (1.2,0.2)};

\addplot[thick] coordinates{(1,0) (2,1)};
\end{axis}

\node[xshift=15.7mm,yshift=-8.3mm,align=center, fill=white] at (plot1.west) {4-QAM};
\node[xshift=-9mm,yshift=-8.3mm,align=center, fill=white] at (plot1.east) {16-QAM};

\end{tikzpicture}%
\label{fig:diffnorm}}
\caption{Information rate per $T_x=1/B_\text{3dB}$ interval: \protect\subref{fig:combinednorm} 4-QAM and 16-QAM  at \SI{25}{\decibel} with different roll-offs $\beta$ and different rates $T_x/T_s$, oversampling $M=4$,  filter length $9$ symbols; \protect\subref{fig:diffnorm} Areas where the respective alphabet is superior. Lower triangle illustrates the faster than Nyquist region.}
\label{fig:newmeasure}
\end{figure*}
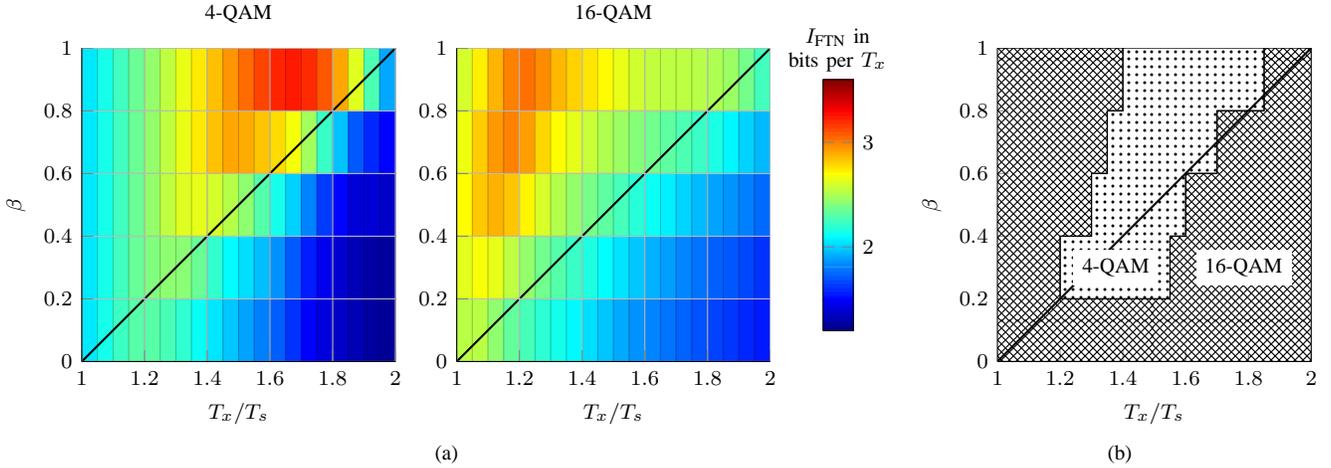

Consider that the information rate $I$ in bpcu from Section \ref{sec:comp} measures the information per symbol duration $T_s$. In the scenario with variable signaling rates, with $T_s \neq T_x$, the effective achievable rate scales with the bandwidth. This can be taken into account by using $I_{\text{3dB}}$ for the faster signaling scenarios given as
\begin{align}
I_{\text{3dB}} & = I \cdot\frac{1}{B_\text{3dB}T_{s}}=I \cdot\frac{T_x}{T_{s}} \text{.} \label{eq:iftn}
\end{align}

Figure \ref{fig:newmeasure} depicts the \SI{25}{\decibel} graphs when applying equation \eqref{eq:iftn} to the results. There is a maximum for each case which is located in the signaling region $T_x/T_s >1$, meaning that the optimal tradeoff can be achieved by using faster than conventional signaling. 4-QAM can benefit more from faster signaling rates due to its property of being more stable when using 1-bit quantization.

The comparison of the results in Figure \ref{fig:diffnorm} shows that for lower signaling values $T_x/T_s$, 16-QAM in combination with oversampling has the higher information rate and that 4-QAM, with or without oversampling, can achieve rates substantially higher than 16-QAM in a certain parameter range.

\section{Conclusion} \label{sec:conc}
In this paper a communications system with matched filters, 1-bit AD-conversion and oversampling has been considered. Gaussian and RRC filters were applied and their influence on the information rate of such a system was investigated. It has been shown that the information rate can be increased compared to using a rectangular pulse with wideband receiver. Furthermore the applicability of faster than Nyquist signaling for this system was explored. The results reveal that a gain in overall system performance is possible depending on the chosen pulse shape, the SNR and the modulation scheme. Furthermore a tradeoff between faster signaling rate and information rate per symbol was considered to find the combination of the two that has the best system performance. This showed that modulation schemes, whose full information rates are more easily accessible for 1-bit quantization, can benefit more from an increased signaling rate. Consequently, choosing the signaling rate wisely is a simple and efficient method to increase the achievable rate, when considering receivers with 1-bit quantization.


\appendix[] \label{appendix}
The considered lower bound on the achievable rate can be calculated based on the entropy rates by
\begin{align}
\lim_{n \to \infty} \frac{1}{n} I(X^n ; \bm{Y}^n) = \lim_{n \to \infty} \frac{1}{n} \left(H(X^n) - H(X^n | \bm{Y}^n)\right) \text{.}
\end{align}
From the i.i.d. assumption follows that $H(X^n)= \sum_{k=1}^n H(X_k)$. Furthermore the equation
\begin{align}
H(X^n | \bm{Y}^n) & \leq \sum_{k=1}^n H(X_k|\bm{Y}^n)	\leq \sum_{k=1}^n H(X_k|\bm{Y}_k)
\end{align}
holds as conditioning can only reduce entropy.




%

\bibliographystyle{IEEEtran}
\bibliography{references}

%
%

\end{document}